\documentclass[twocolumn,prb]{revtex4}

\usepackage{graphicx}
\usepackage{dcolumn}
\usepackage{bm}

\def\dn{{\rm dn}}

\begin{document}

\title{Optical conductivity of one-dimensional doped
Hubbard-Mott insulator}

\author{D. N. Aristov$^\ast$
}
\author{ Vadim V. Cheianov }
\author{A. Luther}
\affiliation{NORDITA,Blegdamsvej 17,DK-2100 Copenhagen, Denmark
}

\date{\today}
\begin{abstract}
We study the optical response of a strongly correlated electron system
near the metal-insulator transition using a mapping to the sine-Gordon
model. With semiclassical quantization,
the spectral weight is distributed between a Drude peak and absorption
lines due to breathers. We calculate the Drude weight, the optical gap,
and the lineshape of breather absorption.
\end{abstract}


\maketitle
Optical absorption is one of the basic tools in studies of the
spectral properties of strongly correlated electron systems
\cite{Ladders,Polymers,Stripes}. An important class of such
systems is comprised of one-dimensional conductors such as
ladder compounds, organic metals and stripes in high $T_c$
cuprates. The electron-electron interaction in
these low-dimensional systems is known \cite{Voit} to result in
spin-charge separation and formation of a Luttinger liquid or
the Mott-Hubbard insulator (MHI). The MHI occurs at commensurate
filling fractions and is characterised by an energy gap in the
charge sector. In this case, optical absorption exhibits a
threshhold at twice the gap energy.  As the Fermi level shifts above the
 gap,
the system conducts, with profound changes to the optical absorption
spectrum.  The optical conductivity summarizes this behavior:
	 $
	\sigma(\omega) =
	 D \  2 \pi   \delta(\omega)+ \sigma_{\rm reg}(\omega),
	 $
with the Drude weight $D=0$ in the insulating phase, and a regular part
$\sigma$ containing contributions from pair excitations and bound
 states,
or excitons \cite{MIT}.

The simplest model where the MHI phase occurs is the Hubbard model with
an on-site repulsion. In the commensurate phase the spectrum of the
 optical
absorption consists of a broad feature lying in the particle-hole
 continuum
\cite{Carmelo00,Jeckelmann00,Controzzi01}. In the more realistic
 extended
Hubbard model, the excitation spectrum generally consists of both the
particle-hole continuum and bound states (excitons)
\cite{Extended,Barford}. Numerical studies
of the extended Hubbard model with nearest- and next-nearest neighbor
repulsions \cite{Essler2001} show the emergence of a sharp exciton peak,
which becomes a dominant feature when the repulsion is
sufficiently strong.

In addition to the numerical analysis of the problem a mapping
to the continuous sine-Gordon (SG) model can be considered \cite{MIT}.
The exciton peaks are seen in the sine-Gordon approach as absorption
on quantum "breather" modes of the model.

Much is known about the commensurate phase because of  the integrable
structure of the corresponding SG theory and the knowledge of its
form-factors due to the Smirnov bootstrap (see \cite{Essler2001} and
 refs
therein). The incommensurate phase is much less studied with analytic
 results
existing for the high frequency expansion of the optical conductivity
in the absence of breathers \cite{MIT}.

In this paper we report our results for the optical conductivity
in the incommensurate phase in the presence of a large number of
breathers, which corresponds to long-range
interactions in the Hubbard chain \cite{Barford}.
It is known that in this limit the SG model
can be analyzed semiclassically \cite{Dashen}. Using the
semiclassical approach we calculate the optical absorption of a
doped MHI as a function of frequency and the carrier density.
Our analytic expression for the Drude
weight (\ref{Eq:Dcla}) is found in an excellent agreement
with a thermodynamic Bethe ansatz calculation.
At high doping the optical weight is contained in the
Drude peak and in a series of well resolved peaks at finite
frequencies, corresponding to excitation of breathers.
We calculate the positions of the peaks
(\ref{Eq:omegan}), and their shape (\ref{Eq:peakshape}). At low
doping peaks merge in a narrow continuous band and
we provide an analytic expression for the optical absorption in
this case. Our results for one spatial dimension agree with
experimental and theoretical evidences  for the peaks in optical
absorption at small energies near the metal-insulator transition in
quasi 1D compounds and in higher dimensions
\cite{Ladders,Stripes,HiD,Fleck,Horsch}.

The optical response of a doped MHI is described by the
sine-Gordon model \cite{MIT,Emery}
        \begin{equation}
	{\cal L}= \frac{1}{2} \left\{(\partial_t \phi)^2 - (\partial_x
	 \phi)^2\right\}
	 +\frac{m^2}{ \beta^2} \cos \beta \phi
	 + h \frac{\beta}{2\pi}\partial_x  \phi
	 \label{Eq:SGLagrangian}
	 \end{equation}
where the  charge and current densities read
        \begin{equation}
	\rho=\frac{\beta}{2 \pi} \partial_x \phi, \quad
	j=-\frac{\beta}{2 \pi}\partial_t\phi.
	\label{Eq:Bosonization}
	\end{equation}
In Eq. (\ref{Eq:SGLagrangian}) $h$ is the chemical
potential controlling the average density
$\langle\rho(x)\rangle=\bar \rho,$ and it is assumed that in
the insulating phase $\bar \rho=0.$
The cosine term in Eq.(\ref{Eq:SGLagrangian}) is related to Umklapp.
The sine-Gordon coupling $\beta$ is determined by the interactions of
the lattice model \cite{Voit}. For free fermions $\beta=\sqrt{ 8 \pi}.$
For infinite $U$ Hubbard model $\beta=\sqrt{4 \pi},$ which corresponds
 to
the refermionization point, where the optical absorption can be found
exactly\cite{Fukuyama}. Below, we will investigate the case of $\beta
 \ll 1$
when the number of breather modes is $\sim 8 \pi / \beta^2$.
he mass $m$ is a characteristic energy scale for a breather excitation
while the particle-hole continuum starts
from energies $\sim 8 m/\beta^2.$

By virtue of (\ref{Eq:Bosonization}) the Kubo
formula for the optical conductivity is written as
        \begin{equation}
	\sigma(\omega)= \left( \frac{\beta}{2 \pi}\right)^2
	{\rm Im}\int_0^\infty d t e^{-i \omega t} G(t),
	\label{Eq:Kubo}
	\end{equation}
where
	\begin{equation} \label{Eq:G}
	G(t)=\int d x
	\langle[\pi(x,t),\phi(0,0)] \rangle.
	\end{equation}
and $\pi= \partial_t \phi $ is the canonically conjugate momentum for
 $\phi,$
such that $[\phi(x), \pi(y)]=i \delta(x-y).$

 It follows from (\ref{Eq:Kubo}) that $\sigma(\omega)$ satisfies a sum
rule
	\begin{equation}
	\int_{-\infty}^{\infty}d\omega\   \sigma(\omega)=
	i\ \frac{\beta^2}{4 \pi}
	\int d x  \langle[\pi(x,0), \phi(0,0)]\rangle=2 \tau,
	\label{Eq:SumRule}
	\end{equation}
with $\tau =\beta^2/8 \pi.$
In the Tomonaga-Luttinger limit, $m=0$ in (\ref{Eq:SGLagrangian}),
the sum rule (\ref{Eq:SumRule}) is saturated by the Drude weight
$ D= {\tau}/ {\pi}.$

In the quasiclassical regime the case $\bar \rho =0$ is very simple.
Eq. (\ref{Eq:SGLagrangian}) reduces to
the Klein-Gordon model, with the ground state $\phi=0.$
The optical conductivity is calculated immediately
\begin{equation}
\sigma_{\rm KG}(\omega)=
\tau \left(\delta(\omega-m)+\delta(\omega+m)\right),
\label{Eq:QC-mott}
\end{equation}

For a finite $\bar \rho$ the quasiclassical ground state found from
the stationary SG equation under constraint
$ \langle \partial_x \phi \rangle = 2 \pi \bar \rho /\beta$
is
	\begin{equation}
	\phi_0(x) = 2\beta^{-1} {\rm am} (2K \bar\rho x )
	\label{Eq:gs}
	\end{equation}
where ${\rm am}(x,k)$ is the Jacobi amplitude function \cite{ArLu}
with the elliptic index $k$. This index is found from the equation
	\begin{equation}
	 m/\bar\rho = 2 k K(k)
	\label{eq-k}
	\end{equation}
with the complete elliptic integral $K$.
Henceforth we will drop the index $k$ in the argument of
the elliptic functions and use
the conventions $k_1 = \sqrt{1-k^2}$, $K(k)=K$, $K(k_1)=K'$,
$E(k)=E$. The standard semiclassical approach requires
the second variation of the action around
the classical solution and an analysis of the spectrum of the resulting
Gaussian action. It was, shown \cite{ArLu} that this
scheme encounters infrared divergences in higher orders of perturbation
theory in $\beta.$ An infrared stable formulation of the theory is
 achieved
by a non-linear change of variables
	\begin{equation}
	\phi = \phi_0
	\left(x +
	\kappa^{-1} \beta\eta \right).
	\label{Eq:change}
	\end{equation}
where $ \kappa=2 \pi\bar\rho.$ The new field variable $\eta(x,t)$ plays
 the role
of the continuous "collective coordinate" \cite{Raja}, which roughly
corresponds
to fluctuations in kinks' positions.

The linearized Lagrangian for the field $\eta$ reads
	\begin{equation}
	{\cal L}=
	\frac12
	w^2(x)
	\{
	(\partial_t\eta)^2  -
	(\partial_x\eta)^2 \}.
	\label{Eq:LiLa}
	\end{equation}
where the weight function, $w$ is defined as
	\begin{equation}
	w(x) = \frac{2K}{\pi} {\rm dn} 2K\bar\rho x.
	\end{equation}
The spectrum of the Lagrangian (\ref{Eq:LiLa}) was analyzed in detail
in \cite{ArLu}. The eigenmodes of (\ref{Eq:LiLa}) are the Bloch
functions
        \begin{equation}
        \eta_q (x, t) = e^{iq x-i \omega_q t} \chi_q(x),
        \label{eta1}
        \end{equation}
where $q$ is the Bloch wave vector and $\chi$ is the modulating Bloch
function with period $1/\bar \rho$.

These solutions are conveniently parametrized by a complex parameter
$\alpha$ as follows. The Bloch wave vector and the eigenfrequency of the
 mode
are given by
        \begin{equation}
        q  = -i 2 \bar \rho K{\rm Z}(\alpha) +
	\pi  \bar \rho,
	\label{Eq:qu} \ \
	\omega =
	2\bar \rho K \,
        {\rm dn}(\alpha),
	\end{equation}
where $Z(\alpha)$ is the Jacobi Zeta function.
The modulating Bloch function is
	\begin{equation}
	 \chi_q(x)= C_\alpha e^{i\pi \bar\rho x} \vartheta_1\left(\pi \bar\rho
 x -
	 \frac{\pi \alpha}
	 {2 K}\right)
	\vartheta_3^{-1}( \pi \bar\rho x).
	\label{Eq:Blom}
	\end{equation}
Here the coefficient $C_\alpha$ is chosen to fulfill the normalization
 condition
	\begin{equation}
	\int dx\, w^2(x)
	\eta_\alpha^\ast (x) \eta_\beta(x)
	=\delta_{\alpha\beta}.
	\end{equation}

The quantized field $\eta(x,t)$ is then represented in the oscillator
basis
        \begin{equation}
        \eta(x,t) = \sum_{\alpha}
        \left(\frac{\eta_{\alpha}(x)}{\sqrt{2\omega_{\alpha}}}
         e^{i\omega_{\alpha}t} b_{\alpha }^\dagger
        + h.c.
        \right)
        \label{Eq:qfield}
        \end{equation}
with operators $b, b^\dagger$ satisfying usual commutation relations
$[b_{\alpha}, b_{\alpha'}^\dagger] = \delta_{\alpha \alpha'}$.

In the limit of small $\beta $ the change of variables (\ref{Eq:change})
 can
be linearized
	$\phi=\phi_0+ w \eta, \ \pi=w \dot \eta$
and from the normal mode expansion (\ref{Eq:qfield}) and the Kubo
formula (\ref{Eq:Kubo}) one obtains for the optical conductivity
	\begin{equation}
	\sigma(\omega)={\tau}
	\sum_\alpha \left|F(\alpha)\right|^2
	( \delta(\omega-\omega_\alpha)+ \delta(\omega+\omega_\alpha)),
	\label{Eq:conductivity}
	\end{equation}
where the form factor
	\begin{equation}
	F(\alpha)= \frac{1}{\sqrt L} \int_0^L d x \eta_\alpha(x) w(x).
	\label{Eq:formfactors}
	\end{equation}

\begin{figure}
	\includegraphics[width=7cm]{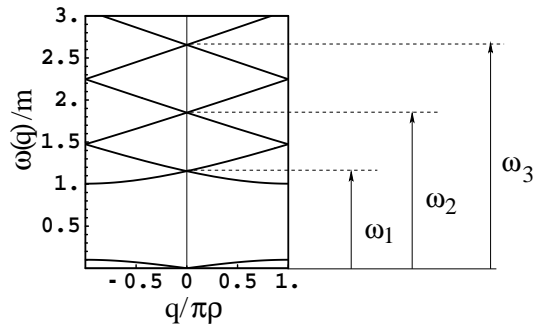}
	\caption{\label{fig:transitions}
	Semiclassical result for the spectrum in the reduced Brillouin zone
	scheme for $\bar \rho/m =0.14.$ The first three allowed transitions
	are labeled
	$\omega_1$,$\omega_2$ and $\omega_3.$
           }
	\end{figure}

The selection rules for the form factors in (\ref{Eq:conductivity})
should allow only transitions between the states of zero wave
vector in the reduced Brillouin zone scheme (see
 Fig.~\ref{fig:transitions}).
This condition combined with (\ref{Eq:qu}) implies that the allowed
indices $\alpha$ and the corresponding frequencies of the
absorption lines satisfy
	\begin{equation}
	2 i K Z(\alpha_n)=\pi(2n-1), \ \
	\omega_n = (m/k) \dn\alpha_n,
	\label{Eq:omegan}
	\end{equation}
where $n=1,2,3,\dots$ and $\alpha_n \in [0, i K').$
For the form factors we find
	\begin{equation}
	|F(\alpha_n)|^2 = \frac{\pi^2}{2K^2}
	\frac{{\rm csch}^{2} \left[\frac{i\pi \alpha_n}{2K}+i K' Z(\alpha_n)
	\right]}{\dn^2\alpha_n- \frac EK}.
	\end{equation}
The lowest frequency $\omega_1$ can be associated with the optical
gap. From Eq. (\ref{Eq:omegan}) (see also Fig.~\ref{fig:transitions})
one finds that $\omega_1$ is always greater than $m$ and is an
increasing function of $\bar\rho.$ One has
$\omega_1 \simeq m$ for $\bar \rho \ll m $ and
$\omega_1 \simeq 2 \pi\bar\rho$ for $\bar \rho \gg m.$
The spacings between the peaks also increase with $\bar\rho,$ from
zero at $\bar\rho\to 0$ to $2 \pi \bar\rho$ for $\bar\rho \gg m.$
The amplitudes $|F(\alpha_n)|^2$ are rapidly decreasing functions
of $n.$ In the high density limit one finds
$|F(\alpha_n)|^2\sim n^{-2} (4\pi\bar\rho/m)^{-4 n}.$

The Drude weight comes from the zero mode $\eta_0$ and is found to be
	\begin{equation}
	D=\frac{\tau}{\pi} \left|F_0 \right|^2 =\frac{\tau \pi}{ 4 E K}
	\label{Eq:Dcla}
	\end{equation}
Combined with definition (\ref{eq-k}) formula (\ref{Eq:Dcla}) gives the
quasiclassical expression for the Drude weight as a function of the
 density
of kinks. The following limiting cases allow for further
 simplifications:

\noindent 1. {\it High density limit.} This limit corresponds to $k\to
 0$
and $K,E \to \pi/2.$ For the Drude weight we get
	$ D\to {\tau}/{ \pi},$
which is the result for the Tomonaga-Luttinger limit.

\noindent 2. {\it Low density limit.} In this limit both $k$ and $E$ are
 close to
unity.
The Drude weight is then given by
	\begin{equation}
	D=\frac {\bar \rho}{2 M_s}.
	\label{Eq:lowden}
	\end{equation}

This result has a simple physical interpretation. In the limit of low
density the kinks form a gas of weakly interacting non-relativistic
particles. In an  external electric field $\varepsilon $ the motion of a
particle is described by  Newton's law $ -i M_s \omega v= \varepsilon $
and the electric current is given  by $ j=-{\bar \rho}\varepsilon/{(i
\omega M_s)}. $ Analytic continuation of this expression gives Eq.
(\ref{Eq:lowden}).

\begin{figure}
	\includegraphics[width=7cm]{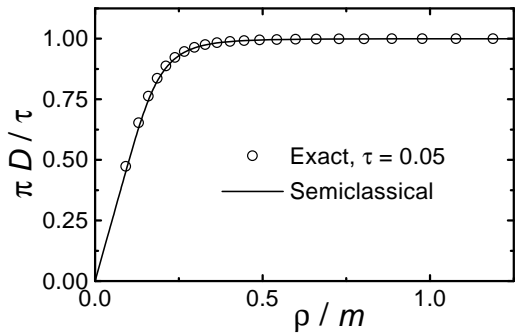}
	\caption{\label{fig:Comparison} Drude weight as a function of soliton
	density in
	the small $\tau $ limit. The solid line corresponds to  the
	semiclassical result,
	Eqs.
	(\ref{eq-k}), (\ref{Eq:Dcla}). The exact values of the Drude weight
	calculated from the Bethe ansatz for $\tau =0.05$ are shown by circles.
           }
	\end{figure}

To check the accuracy of these results we calculate the exact
Drude weight from the thermodynamic Bethe ansatz
\cite{PapaTsve}. A comparison made in Fig.~\ref{fig:Comparison}
shows an excellent agreement between the quasiclassics and the
exact result.

Further analysis needs a different treatment of the cases of
the high and the low density.

In the limit $\rho/m \to 0$ one has $K\to \infty $ so that both the
weight of $\delta-$functions in (\ref{Eq:conductivity}) and the spacings
between them vanish. (In this limit one has
$K'\to \pi/2$, $E\to 1$, $Z(\alpha) \to \tanh \alpha$,
$\dn\alpha \to {\rm sech}\alpha$ and
$\alpha_n \simeq -i (2n+1)\pi \bar\rho/m$,
$\omega_n^2 = m^2 +[(2n+1)\pi\bar\rho]^2 $. ) As a result, the sequence
of $\delta-$functions merges to the continuous function
        \begin{equation}
        \sigma_{\rm reg}(\omega) =
        \frac{\tau \pi\bar\rho }
        {|\omega| \sqrt{ \omega^2-m^2}}
         \frac{\vartheta(\omega^2-m^2)}
         {\sinh^2\frac {\pi}{2m} \sqrt{\omega^2-m^2}}
	\label{Eq:regular}
         \end{equation}
Note the non-integrable singularity
$\sigma(\omega) \sim (\omega - m)^{-3/2}$ at the threshold, which
seemingly violates the sum rule (\ref{Eq:SumRule}). This signals
that the low density limit requires special treatment. A careful
analysis of the fluctuations in this limit \cite{ACL}
leads to the formula
	\begin{equation}
	\nonumber
	\sigma(\omega)=\frac{\tau |\omega|}
	{\pi m\sqrt{\omega^2-m^2}}
	\int_{-\infty}^{\infty} d x
	e^{i x \sqrt{\frac{\omega^2}{m^2}-1} -
	\frac{2 m \bar\rho}{\omega^2}x \coth x},
	\end{equation}
which satisfies the sum rule and coincides with (\ref{Eq:regular})
for  $ \omega-m \agt \bar \rho^2 /m.$

Next we turn to the case of large densities. The series of infinitely
 narrow
peaks in Eq. (\ref{Eq:conductivity}) stems from the
classical picture of the broken translational symmetry of the
ground state Eq. (\ref{Eq:gs}) which is a periodic lattice of
kinks. The true ground state, however, should be
translationally invariant, for arbitrarily small $\beta$.
The symmetry is restored by zero-point fluctuations of the respective
positions of kinks in the lattice. Below we show that such fluctuations
lead to non-perturbative corrections to Eq. (\ref{Eq:conductivity}) and
to the algebraic broadening of the conductance peaks.

In order to analyze finite
$\beta$ corrections to Eq.(\ref{Eq:conductivity})
it is convenient to write the following identity for the correlation
function (\ref{Eq:G})
	\begin{equation}
	\dot G(t)=
	\int d x \frac{m^2}{\beta}
	\langle[\sin \beta \phi(x,t), \phi(0)]
	\rangle,
	\label{Eq:corfeqm}
	\end{equation}
which follows from the equations of motion of the sine-Gordon model.
The optical sum rule is taken care of by imposing the
initial condition
$G(0)=-i.$

With a help of (\ref{Eq:change}) and observing that the ground state
solution $\phi_0$ satisfies the stationary SG equation we write
	\begin{equation}
	 \dot G(t)=\int dx
	\langle[\phi_0''\left(x+ \frac{\beta \eta(x,t)}{\kappa} \right) ,
	\phi_0\left(\frac{\beta \eta(0)}{\kappa}  \right)]
	\rangle_X
	\label{Eq:Gdot1}
	\end{equation}
where the average over the position of the lattice of kinks $X$ is
spelled out explicitly. The function $\phi_0$ is a sum of a
linear and a periodic function of its argument and can be represented
as a Fourier series
	\begin{equation}
	\phi_0\left(x+\frac{\beta\eta}{\kappa}\right)=
	\frac{\kappa x }{\beta}+
	\eta+
	\sum_{l\neq 0} \frac {C_l}{i\beta l}
	e^{i l (\kappa x +\beta \eta)}
	\label{Eq:Fourier}
	\end{equation}
with $
	C_l={\rm sech}
	({\pi l  K'}/{K}).
	$
Using (\ref{Eq:Fourier}) we write Eq. (\ref{Eq:Gdot1}) as
	\begin{equation}
	\dot G(t)=\kappa^2 L^{-1}\int dx d y \sum_l C_l^2
	e^{i \kappa l (x-y)} Q_l(x-y,t),
	\label{Eq:Gdotanswer}
	\end{equation}
where
	\begin{equation}
	Q_l(x-y,t)=\beta^{-2}
	\langle
	[e^{i l \beta \eta(x,t)}, e^{-i l \beta \eta(y)}]
	\rangle
	\label{Eq:Q}
	\end{equation}
In deriving Eq. (\ref{Eq:Gdotanswer}), off-diagonal terms of the form $
\langle[\exp(i m \beta \eta(x,t)),\exp(-i n \beta \eta(0)) ] \rangle$
 for
$ n\neq m$ were dropped, since they vanish in the thermodynamic limit.
The average over the position of the kink lattice $X$ was replaced by
the equivalent integration over $y.$

The limit $\beta \to 0$ in (\ref{Eq:Gdotanswer}) is non-trivial because
of the long-distance singularities characteristic of this problem.
First take this limit na\"{\i}vely by expanding the exponentials in
Eq. (\ref{Eq:Q}) in powers of small $\beta$ and keeping
the leading term. Then
	\begin{equation}
	Q_l(x-y,t)
	\approx
	l^2\langle [\eta(x,t), \eta(y)]
	\rangle.
	\label{Eq:naiive}
	\end{equation}
Using this approximation in formula  (\ref{Eq:Gdotanswer}) immediately
 leads to
the peaked structure of the optical response (\ref{Eq:conductivity}).

Next, consider the average commutator in Eq. (\ref{Eq:Q}) using the
 Gaussian
approximation
	\begin{equation}
	\langle e^{i \beta l \eta(x,t)} e^{-i \beta l \eta(0)}\rangle=
	e^{-\beta^2 l^2 \langle \eta(x,t) \eta(0)\rangle}.
	\label{Eq:Gaussian}
	\end{equation}

Eq. (\ref{Eq:naiive}) is obtained from the first two
terms of the Taylor expansion of the exponential in the right hand side
of (\ref{Eq:Gaussian}). This expansion is accurate while the argument
of the exponential is small, which is not true at large distances,
where $\beta^2 \langle \eta(x,t) \eta(0)\rangle \approx
{\cal K} \log|x^2-v_s^2 t^2|, $ where $\cal K$ is the Luttinger
parameter, \cite{ArLu}. The breakdown of the approximation Eq.
(\ref{Eq:naiive}) at large distances leads to a broadening of the
 conductance
peaks.

A rather lengthy but straightforward analysis, based on Eq.
(\ref{Eq:Gaussian}) shows then that the real part of optical
 conductivity near
the $n$-th peak is given by
	\begin{equation}
	\sigma(\omega_n+\Omega)\propto \sum_l {\rm Re}
	\int_0^{\infty} dt
	\frac{l^2 C_l^2 e^{-i \Omega t}}
	{(\bar\rho^2 (v_n^2-v_s^2)t^2+1)^{{\cal K} l^2 } },
	\end{equation}
where $v_n$ is the group velocity of the bosonic excitations near
 frequency
$\omega_n$, $v_s$ is the sound velocity of bosonic excitations.
We can see, that in the limit ${\cal K}=0$ the delta-function form of
 the
conductivity peak is restored. At any finite value of the Luttinger
 parameter
${\cal K}$ the conductivity will fall off algebraically away from
 $\omega_n$.
	\begin{equation}
	\sigma(\omega_n+\Omega)\propto |\Omega|^{2 {\cal K}-1}
	\label{Eq:peakshape}
	\end{equation}
The width of the peak, on the other hand, is determined by the
 difference of the
sound velocity $v_s$ and the group velocity $v_n$ and is given by
\begin{equation}
\Delta \Omega =\bar\rho \sqrt{v_g^2-v_n^2}
\label{Eq:width}
\end{equation}
In the high density limit one finds
$\Delta \Omega \approx m^2/\rho $
which implies, that peaks of optical conductivity are well
resolved despite the quantum smearing of the lattice structure.

In summary, the optical conductivity of a one-dimensional conductor
 close
to the Hubbard-Mott transition in the quasiclassical limit shows a
 smooth
crossover between the Luttinger liquid and the
Mott-Hubbard insulator. With increasing density the band of breather
absorption splits into a series of well resolved peaks. We  calculated
 the
positions of the peaks (\ref{Eq:omegan}), their shape
 (\ref{Eq:peakshape})
and width (\ref{Eq:width}). Our analytic result for the Drude weight
(\ref{Eq:Dcla}) shows an excellent agreement with the thermodynamic
Bethe ansatz.

We thank A. Rosch and P. Horsch for helpful discussions.

\end{document}